\documentclass[fleqn,usenatbib,usedcolumn]{mnras}

\usepackage{newtxtext,newtxmath}
\usepackage[T1]{fontenc}
\usepackage{ae,aecompl}
\usepackage{graphicx}
\usepackage{amsmath}
\usepackage{amssymb}
\usepackage{color}
\usepackage{times}

\usepackage[3D]{movie15}
\usepackage{hyperref}

\newcommand{\kms}{\mbox{${\rm km\,s}^{-1}$}}
\volume{{\rm in press}}         

\title[Planet-disc kinematic predictions]{Observability of planet-disc interactions in CO kinematics}

\author[S. P\'erez et al.]{Sebasti\'an P\'erez,$^{1,2}$\thanks{E-mail: sperez@das.uchile.cl}
  S. Casassus$^{1,2}$ and P. Ben\'itez-Llambay$^3$
  \\
  $^{1}$Departamento de Astronom\'ia, Universidad de Chile, Casilla
  36-D, Santiago, Chile\\
  $^{2}$Millennium Nucleus on ``Protoplanetary Disks'', Chile\\
  $^{3}$Niels Bohr International Academy,
  Niels Bohr Institute, Blegdamsvej 17, DK-2100 Copenhagen \O{},
  Denmark }

\date{Accepted XXX. Received YYY; in original form ZZZ}
\pubyear{2018}

\begin{document}

\label{firstpage}
\pagerange{\pageref{firstpage}--\pageref{lastpage}}
\maketitle

\begin{abstract}

Empirical evidence of planets in gas-rich circumstellar discs is
required to constrain giant planet formation theories. Here we study
the kinematic patterns which arise from \mbox{planet-disc}
interactions and their observability in CO rotational emission
lines. We perform \mbox{three-dimensional} hydrodynamical simulations
of single giant planets, and predict the emergent intensity field with
radiative transfer.  Pressure gradients at planet-carved gaps, spiral
wakes and vortices bear strong kinematic counterparts. The
iso-velocity contours in the \mbox{CO(2-1)} line centroids $v_\circ$
reveal large-scale perturbations, corresponding to abrupt transitions
from below {\em sub}-Keplerian to {\em super}-Keplerian rotation along
with radial and vertical flows. The increase in line optical depth at
the edge of the gap also modulates $v_\circ$, but this is a mild
effect compared to the dynamical imprint of the planet-disc
interaction. The large-scale deviations from the Keplerian rotation
thus allow the planets to be indirectly detected via the first moment
maps of molecular gas tracers, at ALMA angular resolutions.  The
strength of these deviations depends on the mass of the
perturber. This initial study paves the way to eventually determine
the mass of the planet by comparison with more detailed models.

\end{abstract}


\section{Introduction}

The detection of protoplanets via direct imaging has been notoriously
difficult, requiring extreme high-contrast data and substantial image
filtering. At this point, it is not clear whether the candidate
detections correspond to point-like emission
\citep{2015Natur.527..342S} associated with a circumplanetary disc
(CPD), residual disc features \citep{Tha2016, Fol2017}, or shock
heating locally puffing up the edges of a gap \citep{Hor2017}. These
technological barriers combined with large uncertainties in the
initial conditions of evolutionary tracks \citep[high- or low-entropy
  start models,][]{Spie2012}, hampers the unambiguous detection of
protoplanets and their characterisation. This is a crucial missing
link for understanding the origin of planetary systems.

Dynamical interactions between a gaseous disc and a massive accreting
planet produce structures such as gaps, spiral arms, and vortices
\citep{1980ApJ...241..425G, 1986ApJ...307..395L,
  2007A&A...471.1043D}. Correlations between these structures and the
presence of embedded planets have been studied in dust continuum and
scattered light observations \citep[e.g.][]{2015MNRAS.453L..73D,
  Dong2015, Bae2017}, potentially constraining the location and mass
of the putative planets. However, several other mechanisms account for
these observations, as exemplified by the diversity of theories that
explain the HL Tau ring system \citep{2015MNRAS.453L..78L,
  2016ApJ...821...82O, 2018A&A...609A..50D}.

Density perturbations induced by massive planets also bear kinematic
counterparts. In the presence of a planet embedded in the disc, two
spiral wakes develop in co-rotation with the perturber. In addition,
the gas performs horseshoe orbits where, for a giant planet, a U-turn
is present near the CPD \citep[e.g.][]{Machida2008}. If the viscosity
is low enough, a vortex is also prone to develop at the edge of the
gap \citep[e.g.][]{2001ApJ...551..874L}, which consists in a subtle
modulation on top of the gas Keplerian motion \citep{Dul2013}. Do
these distinct kinematic features bear an observable footprint in the
Doppler maps of molecular lines? Are these signatures within the reach
of current instrumentation, and can they lift the degeneracies in
estimates based on the density field only?

Few works have studied the observability of planet-induced structures
in molecular lines \citep{2015ApJ...807....2C, Fac2017}, and fewer
address their impact on the kinematics.  For example,
\citet{2015A&A...579A.105O} studied planet-carved gaps in line maps
while assuming Keplerian velocities. 

In a previous work we examined the vicinity of the CPD in channel maps
\citep{Perez2015}, i.e. we considered the observability of CPDs by
characterising their morphology and kinematics in CO datacubes using
radiative transfer applied to SPH hydrodynamical simulations, and
subsequent filtering for the $uv$-coverage of the Atacama Large
Millimetre/submillimetre Array (ALMA) Observatory.  We found three
distinct CPD signposts at the vicinity of the protoplanet: (1) compact
emission separated in velocity from the overall Keplerian disc, (2) a
pronounced kink on the Doppler-shifted line emission as it sweeps
across the CPD, and (3) a local increase in the dispersion at the
protoplanet position. The latter allows us to measure the size of the
CPD, which can be used as a proxy for the mass of the protoplanet
\citep{Perez2015} --the radius of a CPD is one third to one half of
the planet's Hill sphere radius \citep[e.g.,][]{Gressel2013}. These
signposts are detectable at the spatial and spectral resolutions
available in CO(2-1) in the extended ALMA array configurations.

Here, we address whether planet-disc interactions can alter the
velocity field to a level that makes the induced perturbations
detectable. To this end, we perform 3D hydrodynamic simulations of
gap-opening planets embedded in locally isothermal discs
(Section~\ref{sec:methods}). These new simulations follow the
evolution of the disc over a longer timescale and with a larger radial
domain than our 2015 calculations, allowing us to study large scale
structures (Section~\ref{sec:hydro}).  We combine these simulations
with 3D Monte Carlo radiative transfer to calculate synthetic velocity
maps (first moments) of the $^{12}$CO line (Section~\ref{sec:rt}). The
outcome of the simulations and the predictions are then probed for
kinematic signposts inherent to planet-disc interactions
(Section~\ref{sec:results}). Implications are summarised in
Section~\ref{sec:discussion}.

\section{Methods}
\label{sec:methods}

\subsection{3D hydrodynamic simulations}
\label{sec:hydro} 

\begin{figure*}
\includegraphics[width=0.82\textwidth]{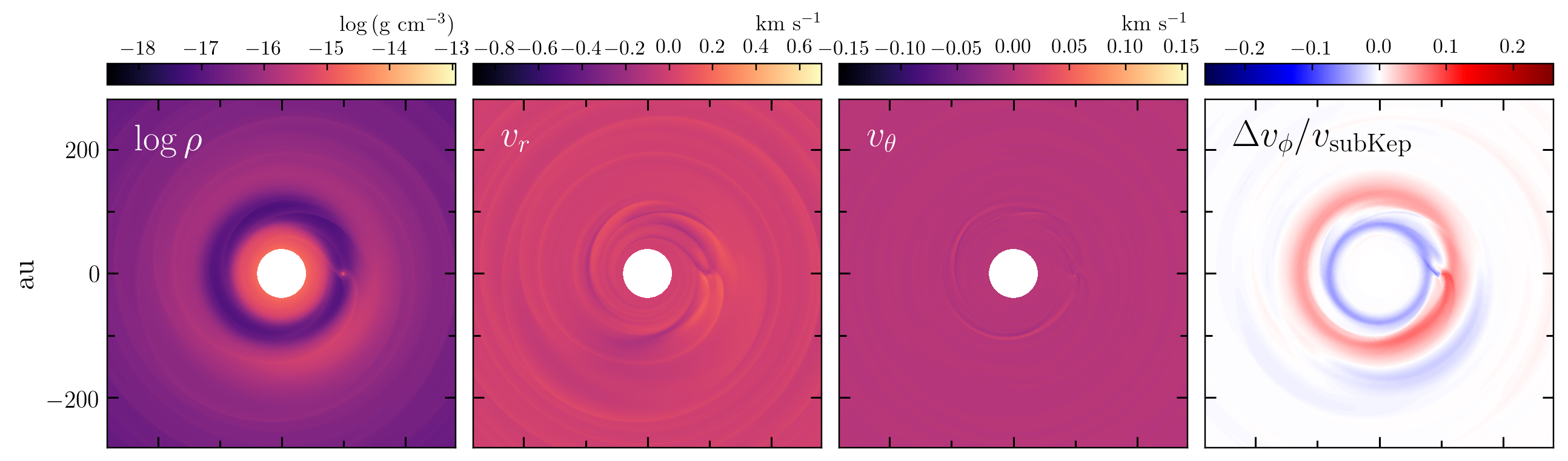} 
\includegraphics[width=0.82\textwidth]{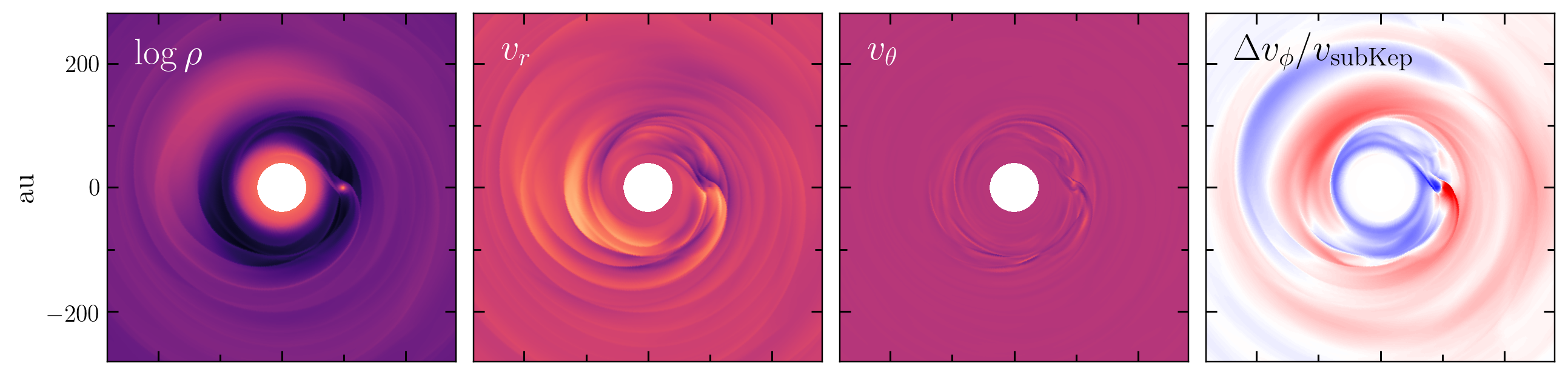} 
\includegraphics[width=0.82\textwidth]{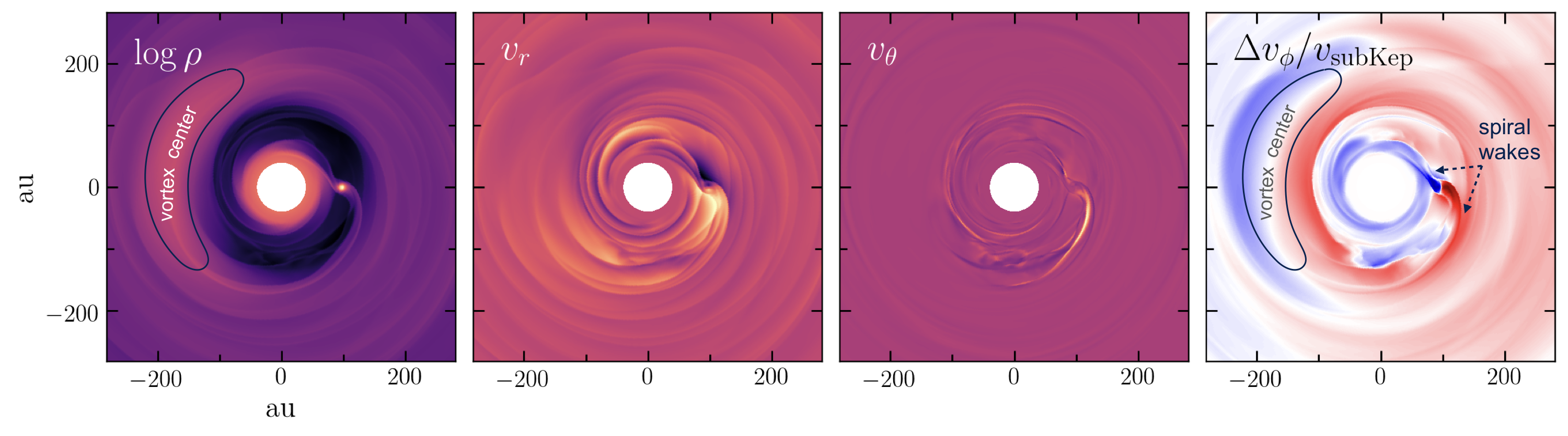} 
  \caption{Simulation results for {\tt P01J} (1~M$_{\rm Jup}$, top
    row), {\tt P05J} (5~M$_{\rm Jup}$, middle) and {\tt P10J}
    (10~M$_{\rm Jup}$, bottom) in the midplane after 1000 orbits.
    Left to right: logarithm of the gas density ($\rho$), radial
    velocity ($v_r$), colatitude velocity ($v_\theta$), and the
    deviations from the sub-Keplerian equilibrium solution in the
    azimuthal component $\Delta v_\phi/v_{\rm subKep}$.  Values are in
    physical units.  Simulations are scaled to $r_{\rm
      p}=100$\,au. The gaps, spiral wakes, and vortices are prominent
    in all three simulations. In $v_\theta$, positive values
    represents gas moving towards the midplane, while $v_r > 0$
    represents gas moving outwards. The disc rotation is
    counter-clockwise. The vortex and spiral wakes are labelled in the
    bottom right panel. }
\label{fig:sims}
\end{figure*}

The hydrodynamical evolution of a gaseous disc with an embedded giant
planet is simulated using the FARGO3D code \citep[Version
  1.2,][]{2016ApJS..223...11B}, with the FARGO algorithm activated
\citep{2000A&AS..141..165M}. Fluid equations are solved in a spherical
grid, where radius, azimuth and colatitude are denoted by $r$, $\phi$,
and $\theta$, respectively.

The equation of state is chosen as locally isothermal. This
simplifying assumption, although far from being a realistic
approximation, is enough to produce the planet-disc interactions
structures we aim to study in the kinematics, and even form a CPD
around the planet \citep{Gressel2013}.

A planet is fixed on a circular orbit of radius $r_{\rm p}$. The grid
domain spans from 0 to 2$\pi$ in $\phi$, $0.4r_{\rm p}$ to $4 r_{\rm
  p}$ in $r$, and $\pi/2-0.3$ to $\pi/2$ (the disc midplane) in
$\theta$, covering at least three vertical scale heights ($H = h/r$)
within the domain. The mesh is evenly spaced in all directions using
$(n_r, n_\phi, n_\theta) = (400\times600\times40)$ grid cells.  The
planet's potential is softened over a scale $0.3H$, which is of the
order of the size of a CPD around a Jupiter-like planet at 100~au.
The mass of the planet is gradually increased over the first few
orbits using a tapering function.

The density and azimuthal velocity fields are extrapolated at the
radial boundaries \citep{2016ApJS..223...11B}, with reflecting
boundary conditions at the midplane. Damping buffer zones near the
radial boundaries of the mesh minimize spurious reflections
\citep{2006MNRAS.370..529D}. The units of the calculations are such
that the gravitational constant, the stellar mass, and the size of the
planet's orbit, $r_{\rm p}$, are all equal to one.

We adopt an $\alpha$-viscous model for the disk, and set
$\alpha=10^{-4}$. This $\alpha$ value allows vortices produced at the
gap edge to be sustained for thousands of orbits \citep{Fu2014}. The
aspect ratio of the disc is set to $h = h_0 (r/r_{\rm p})^f$, with
$h_0 = 0.08$ and $f=0.15$.  The initial surface density is $\Sigma =
\Sigma_0(r/r_{\rm p})^{-1}$, with $\Sigma_0 = 0.09$~g\,cm$^{-2}$ at
$r_{\rm p} = 100$~au (total disc mass is
7$\times$10$^{-4}\,M_\odot$). We adopt a 1\,M$_\odot$ star.  As the
disc is not self-gravitating and does not exerts back-reaction onto
the planet, the value of $\Sigma_0$ does not have any impact on the
dynamics of the system.  However, we note that it does have an effect
on the optical depth of the emission line. This is addressed towards
the end of Section~\ref{sec:results}.

The initial condition corresponds to the steady-state solution
assuming hydrostatic equilibrium in the vertical direction
\citep[see][Apendix A]{Masset2016}. We then carry out three different
simulations, each starts from the same initial condition, including a
giant planet of 1, 5, or 10 Jupiter masses, labeled {\tt P01J}, {\tt
  P05J} and {\tt P10J}, respectively.  Further disc properties are
presented in physical units in Section~\ref{sec:rt}.  Throughout the
paper, whenever time is given in ``orbits'', it refers to the orbital
period at the planet's location, i.e. about 1000~yr for $r_{\rm
  p}$=$100$~au.

These simulations were performed using the parallel Graphics
Processing Unit (GPU) mode of FARGO3D. GPU acceleration allows full 3D
hydrodynamic simulations of large discs to be evolved over long
timescales. A GPU cluster with four Tesla K80 HPC cards with enabled
Peer-to-Peer communications between GPUs \citep{2016ApJS..223...11B},
reached an average performance of $\sim$20\,orbits/hr for the
simulations described here.

\subsection{3D radiative transfer}
\label{sec:rt}

Our choice of observables are spectrally- and spatially-resolved
moment maps which are calculated from synthetic CO channel
maps. Centred at 230.538~GHz, the $^{12}$CO $J$=$2$-$1$ emission line
is bright and routinely observed by millimeter interferometers (it
falls within ALMA's Band-6). We compute these synthetic images using
the radiative transfer code {\sc radmc-3d} \citep[version 0.41,
][]{2012ascl.soft02015D}. We note that these predictions are valid for
similar transitions (e.g., $^{12}$CO $J=3-2$ or $J=6-5$) in LTE. The
density and velocity fields obtained from the hydrodynamical
simulations are first scaled to $r_{\rm p}=100$\,au and then input in
{\sc radmc-3d}, preserving their grid structure and resolution.
 
A 7000\,K star with 1\,R$_\odot$ is placed at the centre of the
grid. In addition, a mock dust distribution consisting of a mix of
30\% carbon and 70\% silicates is added following the simulated gas
density. This dust distribution is only used to perform the
temperature calculation via the Monte Carlo {\sc radmc-3d} module {\tt
  mctherm}. Studying the continuum emission arising from dust thermal
radiation requires a proper treatment of dust coupling through multi
fluid simulations \citep[][Baruteau et al. {\it in prep}]{Bar2016b}
and is beyond the scope of this paper. We caution that our predictions
do not take into account optical depth effects due to dust opacities.

Line radiative transfer is done in non-LTE with the Large Velocity
Gradient (Sobolev) mode in {\sc radmc-3d}. The molecular data is from
the LAMDA database. Line widths include thermal broadening and a local
(spatially unresolved) microscopic turbulence set to a constant value
of 0.1\,km\,s$^{-1}$.  The $^{12}$CO abundance relative to molecular
hydrogen is assumed to be the standard value 10$^{-4}$. Our nominal
simulated discs are inclined by 20$^\circ$ around a PA of 140$^\circ$
(East of North). The disc near side is the South-West and its rotation
is counter-clockwise.

The resulting synthetic cubes have a total bandwidth of
6\,km\,s$^{-1}$, with 120 individual channels of 0.05\,km\,s$^{-1}$
each.  These synthetic data cubes represent our sky model which is
subsequently convolved with a 50\,mas circular beam (a standard
angular resolution in ALMA Band 6 observations).  Our observables
correspond to the intensity-weighted velocity centroid of the
Doppler-shifted line emission (first moment maps).  Since no dust is
considered at the ray-tracing stage, the synthetic line emission cubes
do not suffer from continuum over subtraction due to CO absorption
\citep[see][their fig. 8]{Boe2017}.

To address the observability of the planet-disc interaction signatures
under realistic observing conditions, we first scale our $^{12}$CO
emission maps to an average flux density of 30~mJy~beam$^{-1}$ (in a
50~mas beam) at 100~au \citep[similar to the flux levels reported for
  TW Hya,][]{2018ApJ...852..122H}, then we input these into CASA 5.1
{\tt simobserve}.  A 10\,h integration in configuration C43-9
(13.8\,km baseline) combined with a 3\,h integration in C43-5 (1.4\,km
baseline) is needed to produce a well-sampled $uv$-coverage that
unambiguously recovers the signatures (calibrations will require a
similar amount of time).  We further corrupted the visibilities with
the task {\tt ms.corrupt()} to obtained an RMS level in each
0.25~km~s$^{-1}$ channel of $\sim$1~mJy~beam$^{-1}$. The beam is
$\sim$50$\times$80~mas beam (Briggs weighting 0.5). The
signal-to-noise achieved at the planet's radius is $\sim$30.  The low
inclination of our synthetic disc ($i$=20$^{\circ}$) implies that a
high spectral resolution (0.25~km~s$^{-1}$) is required to resolve the
kinematics.  The exact observing times and configurations required
will depend on specific properties of the target. For example, for
higher inclination angles the kinematics can be traced with lower
spectral resolution.  Finally, the first moment maps were computed in
the CLEANed spectral cubes (produced with CASA {\tt tclean}) by
performing a gaussian fit. The centroids of these gaussians comprise
the velocity fields shown in the fourth column in Fig.~\ref{fig:mom1}.

The temperature field obtained via the Monte Carlo solution of thermal
equilibrium yields similar values to the isothermal power law used in
the hydro simulation. The temperatures in our RT domain within 200~au
are well above 25~K, preventing CO freeze-out from being an issue.
The sound speed at the location of the planet is of the order of
$\sim$90~m~s$^{-1}$, which is small compared to the deviations in
Keplerian motion which we present in Section~\ref{sec:results}.
Heating from the planet is not considered in our calculation --its
impact would be at small scales near the CPD. Changes in the
thermodynamics of the whole disc, e.g an adiabatic disc where
compressional heating from spiral wakes becomes relevant, could in
principle produce distinct kinematic signatures. Exploring these
thermodynamic effects is beyond the scope of our paper.

\section{Results}
\label{sec:results}

Midplane densities after 1000 orbits are shown in the left column of
Fig.~\ref{fig:sims}. The gap edge is Rossby unstable, leading to the
formation of a large-scale vortex in the outer disc
\citep{Lovelace1999, Var2006}. The vortex is stable and long-lived due
to our choice of low viscosity \citep{Fu2014}.  After 1000 orbits,
well-defined gaps, spiral wakes in the inner and outer disc, as well
as a lopsided overdensity region in the outer disc, are observed in
all our simulations.  Detailed descriptions of these structures as
density features are found elsewhere in the literature
\citep[e.g.,][]{Fu2014, Bae2017, Ham2017}.

In the absence of a perturber, the disc follows a nearly pure
Keplerian motion $v_{\rm subKep}$, with $v_r$ and $v_\theta$ values
that are negligible compared to the local azimuthal velocity. An
observation of such unperturbed disc, inclined w.r.t the
line-of-sight, displays a velocity field whose iso-velocity contours
follow a characteristic 'dipole pattern'. This Keplerian pattern is
symmetric w.r.t a straight line at systemic velocity, parallel to the
disc's minor axis.

\begin{figure}
  \centering\includegraphics[width=0.95\columnwidth]{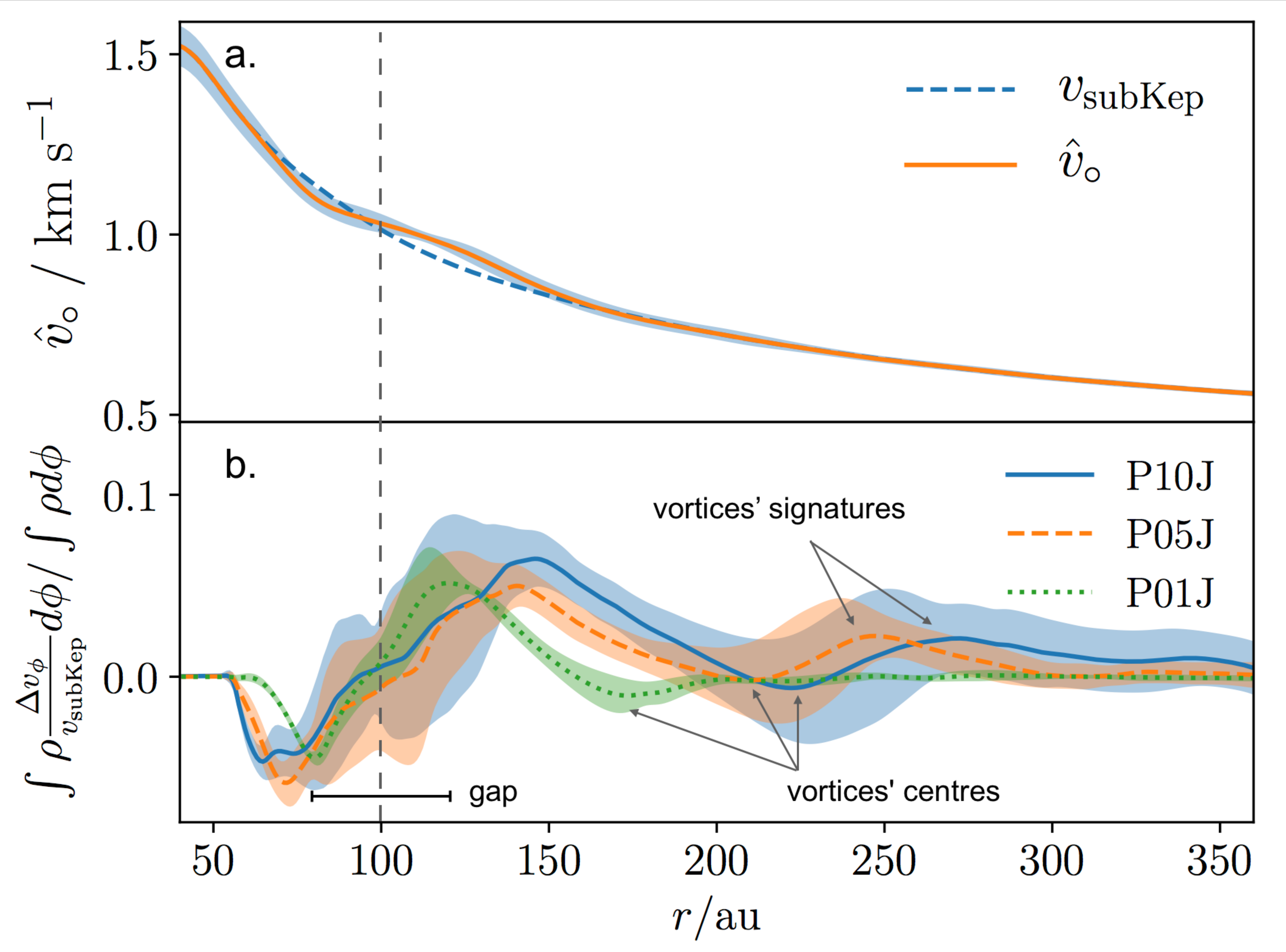}
  \caption{(a) Average Keplerian profile $\hat{v}_\circ$ calculated
    from the synthetic moment $v_\circ$ in Fig.~\ref{fig:mom1}. (b)
    Density-weighted average along the azimuthal axis for $\Delta
    v_\phi/v_{\rm subKep}$ (see Fig.~\ref{fig:sims}), as a function of
    radius. The profiles' standard deviations are shown as a shaded
    area around the average values.}
\label{fig:profiles}
\end{figure}

\begin{figure*}
  \includemovie[controls=false, repeat=100,
    text={\includegraphics[width=.9\linewidth]{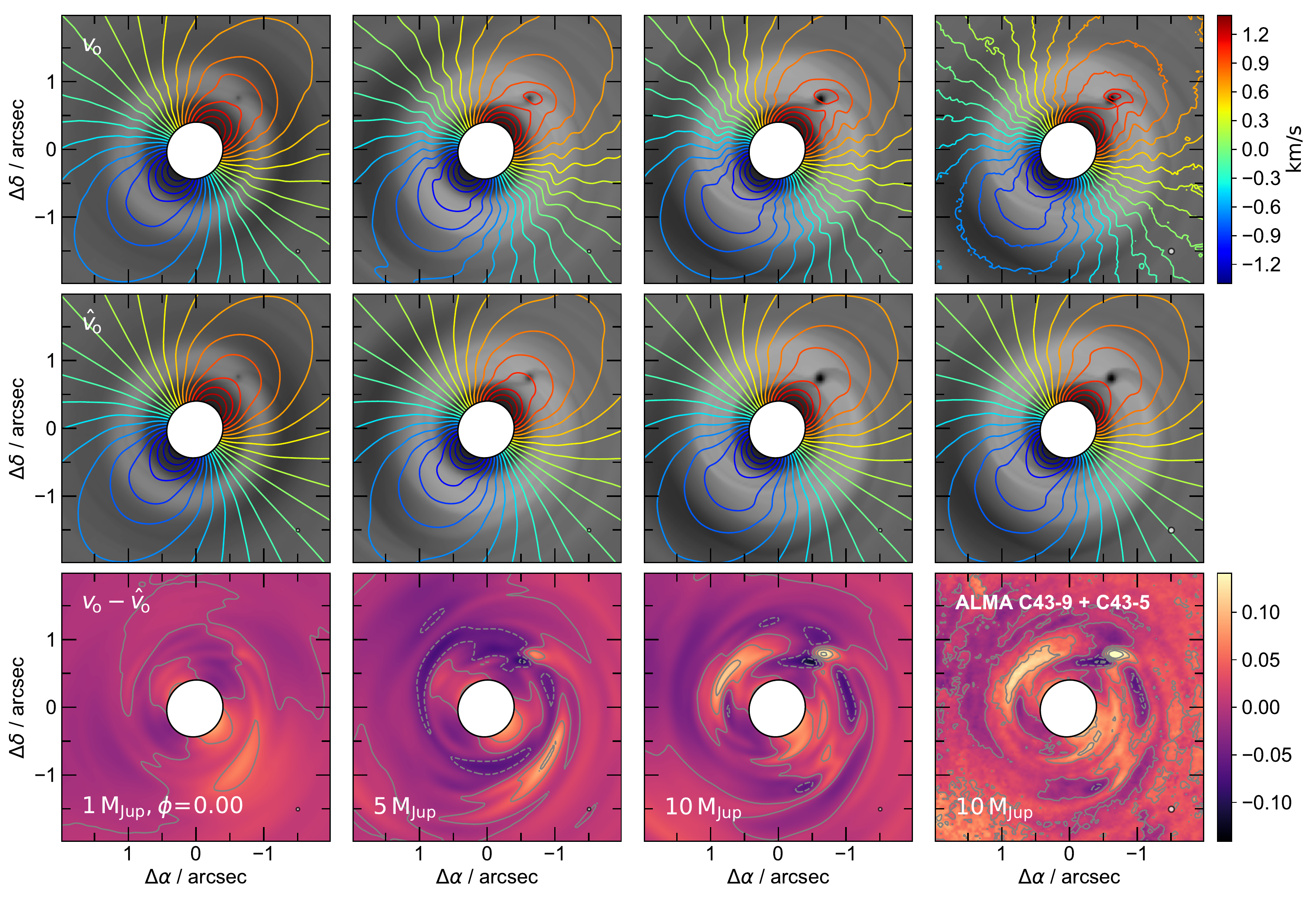}}]{}{}{anc/movie_m1_avg_simobserve_12co21_Rs100.mov}
  \caption{ Upper panels show the first moment map ($v_\circ$)
    predictions in $^{12}$CO(2-1) with embedded companions of (from
    left to right) 1, 5, and 10~M$_{\rm Jup}$, after 1000 orbits
    ($\sim$0.5~Myr) and convolved with a 50\,mas beam. The fourth
    column (right) shows the prediction for the 10~M$_{\rm Jup}$ case
    after filtering through ALMA Cycle 6's uv-coverage (configuration
    C43-9 plus C43-5) with an RMS noise of $\sim$1~mJy~beam${^{-1}}$
    (see Section~\ref{sec:rt}).  Click to play a short movie showing
    how the first moments change with the location of the planet ({\sc
      adobe reader} recommended). The first frame shows the planet
    along the disc PA. The underlying gray color maps show the dust
    continuum following the same density as the gas. These continuum
    maps are only used to highlight the positions of the companions
    and vortices.  Middle panels show $\hat{v}_\circ$ computed
    directly from the $v_\circ$ (see Section~\ref{sec:results}), which
    serve as a proxy for the unperturbed Keplerian disc profile.
    Contours show the projected velocities with colours ranging from
    -1.4 (blue) to 1.4~km~s$^{-1}$ (red), in steps of
    0.15~km~s$^{-1}$. The bottom panels show the deviation from
    Keplerianity, i.e., the difference between top and middle
    panels. The $x$ and $y$ axes show angular offset from the stellar
    position, in arc-second. }
  \label{fig:mom1}
\end{figure*}

Fig.~\ref{fig:sims} and ~\ref{fig:profiles} allow us to study the
kinematics of gaps, spiral wakes, and vortices, while
Fig.~\ref{fig:mom1} shows the observable counterparts of these
structures.  Fig.~\ref{fig:sims} shows the evolved velocity fields
$v_r$ and $v_\theta$ (middle panels), and the relative deviations from
Keplerian rotation \mbox{$(v_\phi-v_{\rm subKep})/v_{\rm subKep}$}
(right panels) after 1000 orbits. Radial profiles of the azimuthal
deviations are shown in Fig.~\ref{fig:profiles}. In turn,
Fig.~\ref{fig:mom1} shows iso-velocity contours in the predicted
$^{12}$CO(2-1) line centroids $v_\circ$.  The largest deviations occur
close to the planet, near the CPD.  The observability of this CPD
signature requires observations with a resolution comparable to the
size of the CPD ($\sim$1/3 Hill radius) and is described in
\citet{Perez2015}.

Relative deviations $\Delta v_\phi/v_{\rm subKep}$ from Keplerian
motion are notorious in the gap for all simulations.  As the gas in
the outer wall speeds up, the inner wall slows down producing zones
where gas transitions from below sub-Keplerian to {\em super}
Keplerian.  For large planet masses ({\tt P05J} and {\tt P10J}), $v_r$
and $v_\theta$ values can grow up to $\sim$20\% of the local azimuthal
speed, while for 1\,M$_{\rm Jup}$ these values increase only up to
$1-5$\% the local azimuthal speed ($\sim$3\,km\,s$^{-1}$ at 100\,au).
These perturbations are induced by gravitational forces exerted by the
planet as well as pressure and viscous forces. In our isothermal
simulations, the pressure field is proportional to the density field
(left panels in Fig.~\ref{fig:sims}) multiplied by $1/r$.

Across the gap (between $\sim$80 and $\sim$120~au), there is a
gradient in the azimuthal average of $\Delta v_\phi/v_{\rm
  subKep}$. Here the mean deviation increases monotonically from -5 to
+5\% (Fig.~\ref{fig:profiles}b). This is shared by all our
simulations.  At 100~au, this gradient translates into a
$\sim$0.3~\kms deviation which is indeed observable, but its magnitude
does not depend strongly on the planet's mass.


The maximum departure from Keplerianity occurs at the outer spiral
wake, where a strong dependency with the mass of the planet is
observed, with maximum deviations of 26\%, 18\% and 9\% for {\tt
  P01J}, {\tt P05J}, and {\tt P10J}, respectively.  This is
demonstrated in Fig.~\ref{fig:profiles}a. $v_\theta$ traces the
departure from hydrostatic equilibrium which, as expected, is stronger
at the spiral wake where material streams in from the poles onto the
CPD.


The vortex signature appears in $\Delta v_\phi/v_{\rm subKep}$ as a
subtle motion on top of the Keplerian disc. The side of the vortex
which is closest to the gap is faster than $v_{\rm subKep}$, while the
outer part is slower.  The strength of this super Keplerian region
depends on the perturber's mass. Fig.~\ref{fig:profiles}b shows that
at radii larger than the vortex centre the peak Keplerian deviations
attain up to 5\% for {\tt P05J} and {\tt P10J}, but negligible (<1\%)
for {\tt P01J}.  The bottom panel displays that these deviations are
significant in the azimuthal average as well, with up to 2-3\%
deviation over the outer disc.


The first moment prediction for our three simulations are shown in
Fig.~\ref{fig:mom1}. The iso-velocity contours in the line centroids
$v_\circ$ show wiggling substructure over various areas in the disc,
becoming ubiquitous and strong for {\tt P05J} and {\tt P10J}. These
wiggling structures correspond to parcels of gas that experience
pressure gradients that make them attain values from below
sub-Keplerian to super Keplerian motion (see $\Delta v_\phi/v_{\rm
  subKep}$ maps in Fig.~\ref{fig:sims}). These pressure gradients are
mainly due to gravitational interactions with the embedded planet. The
gas within the gap experiences such pressure gradients, leading to a
sawtooth signature in the iso-velocity contours at all azimuths.

In first approximation the wiggles in the sky map $v_\circ(\vec{x})$
are perturbations on an approximately Keplerian background flow,
$\hat{v}_{\circ}(\vec{x})$.  This background flow includes modulations
from the line optical depth $\tau(\vec{x})$, especially at the outer
gap edge. We use $v_\circ(\vec{x})$ directly to estimate
$\hat{v}_{\circ}(\vec{x})$, as would be done for an actual
observation.  We deproject $v_\circ(\vec{x})$ by the disk inclination,
\citep[e.g. as in][]{2018MNRAS.tmp..868C}, transform to polar
coordinates about the star, and fit $v_\circ(r, \phi)$ with a harmonic
modulation of the form $\hat{v}_{\circ}(r,\phi) =
\hat{v}(r)\cos(\phi)$.  A projection back to the sky plane gives
$\hat{v}_{\circ}(\vec{x})$. The profile $\hat{v}_\circ$ is shown in
Fig.~\ref{fig:profiles}a, where it is compared against a prediction
$v_{\rm subKep}(r)$ calculated in the same way as $\hat{v}_\circ$ but
from a prediction where the gas is forced to follow pure sub-Keplerian
motion.  We see that \mbox{$|(v_{\rm subKep}(r)-\hat{v}(r))| \lesssim
  \hat{\sigma}(r)$}, where $\hat{\sigma}(r)$ is the standard deviation
of the perturbations $\sqrt{\langle (v_\circ - \hat{v})^2
  \rangle_\phi} $ along azimuth, confirming that $\hat{v}(r)$ is a
good proxy for the unperturbed flow. We see from Fig.~\ref{fig:mom1}
that the sharp rise in $\tau(\vec{x})$ at the outer gap edge results
in a shallow wiggle of $v_\circ(\vec{x})$, much smaller than the
perturbations over $\hat{v}(\vec{x})$.  We note that, for discs with
larger inclinations and significant flaring, a conical transformation
would work better than the simple polar mapping we apply here.

The difference between $\hat{v}_\circ(\vec{x})$ and
${v}_\circ(\vec{x})$ allows us to remove optical depth effects and
quantify the strength of the dynamical influence of a planet (lower
panels in Fig.~\ref{fig:mom1}). Deviations solely due to planet-disc
interactions range from -0.15 to +0.15~km\,s$^{-1}$ over spatial
scales comparable to the gap width and the azimuthal expanse of the
trailing wake and vortex. 

We identify the strongest wiggles near the outer spiral wake and in
front of the CPD (our mock disc rotates counter-clockwise). This is
quantified in the bottom panels of Fig.~\ref{fig:mom1}, which show the
magnitude of these deviations after the subtraction of a Keplerian
first moment map. The maximal deviation is seen when the planet is
near the disc's minor axis. This signature decreases when the planet
approaches the disc PA (see animation in Fig.~\ref{fig:mom1}). Upon
crossing the disc minor axis, the sign of the deviation changes. When
a parcel of super Keplerian gas is on the redshifted (blueshifted)
half of the disc, it will be redder (bluer) than Keplerian, pulling
the iso-velocity contours ahead. Gas moving below Keplerian speed has
a signature in the opposite direction. The outer spiral wake
corresponds to gas at super Keplerian speed and its signature appears
as a sawtooth or kink in $v_\circ$. {\tt P05J} and {\tt P10J} develop
secondary and tertiary spiral arms in the inner disc which also bear a
kink in the iso-velocity contours.

\section{Summary}
\label{sec:discussion}

We performed 3D hydrodynamic simulations to study the dynamical impact
of a planet in the velocity field of circumstellar gas.  Kinematical
counterparts in the three components of the velocity field were
identified for the planet's spiral wakes, the gap, and the outer disc
vortex. The observability of these kinematic structures was then
studied by computing first moment maps of $^{12}$CO line emission
using 3D radiative transfer.

Discs with embedded planets produce deviations from Keplerian
kinematics which are observable at ALMA resolutions. The magnitude of
these deviations depend on the mass of the planet and they are most
notorious at the locations of the spiral wakes, across the gap, and at
the inner secondary (and tertiary) arms.

We conclude that evidence for planet-disc interaction can be detected
by probing for non Keplerian motion in the velocity field of nearby
discs.  In principle the measurement of the corresponding protoplanet
masses can be achieved by comparison with hydrodynamical simulations,
further work is needed to assess model biases with the inclusion of
radiation transport, planet heating and the impact of multiple and/or
eccentric planets. From this initial study, we anticipate that 50~mas
resolutions are sufficient to pick up the kinematic signal of a
$\geq$1\,M$_{\rm Jup}$ planet at 1~arcsec (100~au) around a
1\,M$_\odot$ star.

During the review process of this Letter two new articles reported the
kinematic detection of protoplanets in molecular line observations of
HD~163296.  \citet{Pinte2018} observed the kink in the isovelocity
maps due to an embedded planet, as in the generic predictions by
\citet{Perez2015}, thus identifying the presence of a giant planet at
260~au. The planet-to-star mass ratio puts this detection in a similar
regime as our new 1~M$_{\rm Jup}$ simulation, but the line dataset
used by \citet{Pinte2018} is not deep enough to produce a first moment
map with which to test for the large-scale signatures that we report
here. While their simulations indeed reproduce the localised signature
(i.e. the kink), the large-scale kinematic structures are not present,
which may be due to the short timespan of their SPH calculation. At
the same time, \citet{Teague2018} detected the radial modulation in
the averaged azimuthal velocity imparted by the pressure gradient in
two gaps, at 83 and 137 au, which can be explained by embedded
protoplanets.

In this new Letter, we have improved our 2015 predictions with
simulations over long time spans which have reached steady state,
thanks to which we find that the kinematic perturbations from
planet-disc interactions extend over large scales in the disk.  We
propose that this large scale structure can be used to distinguish the
origin of the velocity perturbations, and eventually pin-point the
perturbers where the deviations from Keplerian rotation changes sign.

\section*{Acknowledgements}

We thank the anonymous reviewer for helpful comments. We acknowledge
support from the government of Chile grants Millennium Scientific
Initiative RC130007, CONICYT-Gemini 32130007, and FONDECYT 1171624.
These simulations were run with the Brelka cluster, hosted at
DAS/U. de Chile (Fondequip EQM140101). This project has received
funding from the European Union's Horizon 2020 research and innovation
programme under grant agreement No 748544 (PBLL).

\bibliographystyle{mnras}

\bsp
\label{lastpage}

\end{document}